%% file: main.tex
\documentclass{article}

\usepackage[emptyfooter,nonatbib]{neurips_2025}

\usepackage[utf8]{inputenc} 
\usepackage[T1]{fontenc}    
\usepackage{hyperref}       
\usepackage{url}            
\usepackage{booktabs}       
\usepackage{amsfonts}       
\usepackage{nicefrac}       
\usepackage{microtype}      
\usepackage{xcolor}         

\usepackage{listings}

\title{Dataset-Agnostic Recommender Systems}

\input{my_authors}

\begin{document}

\maketitle

\input{my_content}
\appendix

\bibliographystyle{plain} 
\bibliography{biblio} 


\end{document}

%% file: my_authors.tex
\author{%
  Tri Kurniawan Wijaya\\ 
  Huawei Ireland Research Centre \\
  Dublin, Ireland\\
  \texttt{tri.kurniawan.wijaya@huawei.com}
  \And
  Edoardo D'Amico\\ 
  Huawei Ireland Research Centre \\
  Dublin, Ireland\\
  \texttt{edoardo.damico@huawei-partners.com}
  \And
  Xinyang Shao\\ 
  Huawei Ireland Research Centre \\
  Dublin, Ireland\\
  \texttt{xinyang.shao@huawei-partners.com}
}

%% file: my_content.tex
\begin{abstract}
Recommender systems have become a cornerstone of personalized user experiences, yet their development typically involves significant manual intervention, including dataset-specific feature engineering, hyperparameter tuning, and configuration. To this end, we introduce a novel paradigm: Dataset-Agnostic Recommender Systems (DAReS) that aims to enable a single codebase to autonomously adapt to various datasets without the need for fine-tuning, for a given recommender system task. Central to this approach is the Dataset Description Language (DsDL), 
a structured format that provides metadata about 
the dataset's features and labels, 
and allow the system to understand dataset's characteristics, 
allowing it to autonomously manage processes like feature selection, missing values imputation, noise removal, and hyperparameter optimization. By reducing the need for domain-specific expertise and manual adjustments, DAReS offers a more efficient and scalable solution for building recommender systems across diverse application domains.
It addresses critical challenges in the field, such as reusability, reproducibility, and accessibility for non-expert users or entry-level researchers. 
%
%
With DAReS, we hope to spark community's attention in making recommender systems more adaptable, reproducible, and usable, with little to no configuration required from (possibly non-expert or entry-level) users.
\end{abstract}

\section{Introduction}
\input{content_intro}

\section{Dataset-Agnostic Recommender Systems (DAReS)}
\label{sec:dares}
\input{content_dares}

\subsection{Comparison to Traditional Recommender Systems}
\label{sec:comparison}
\input{content_comparison}

\section{Automation Levels of Recommender Systems}
\label{sec:leveling}
\input{content_leveling}

\section{Conclusion and Future Work}
\label{sec:conclusion}
\input{content_conclusion}

%% file: content_intro.tex
Recommender systems are essential in delivering personalized content across industries such as e-commerce and streaming services \cite{ricci2010introduction}. Traditional recommender systems, however, require significant manual configuration and domain expertise to adapt to new datasets, limiting their scalability and
reusability~\cite{recwild_2023, hidasi2023effect, automl_position, liu2022online, autorec}. 
These manual processes, including feature engineering, model selection, and hyperparameter tuning, often make it challenging to deploy and reproduce effective models 
consistently~\cite{repr_dacrema,are_dacrema_2019,shao2024rboard}.

To address these challenges, we introduce the \textit{Dataset-Agnostic Recommender System (DAReS)}, which eliminates the need for dataset-specific configurations. DAReS uses the \textit{Dataset Description Language (DsDL)} to describe the key properties of any dataset, allowing fully autonomous feature engineering, model selection, and hyperparameter tuning. This framework enables high-quality recommendation models to be generated with minimal human intervention, making advanced recommendation technologies more accessible.

The key innovation of DAReS is its ability to function as a \textit{zero-configuration} system. Unlike foundational models in NLP (Large Language Models) \cite{brown2020language,raffel2020exploring,touvron2023llama}, 
foundational recommender models seems to be impractical due to variability in dataset 
features~\cite{cantador2011second, wu2023mmgef}. 
DAReS instead leverages DsDL to provide context, allowing it to automatically determine suitable preprocessing, feature transformations, and model configurations.
Despite its advantages in adaptability, reusability, and automation, DAReS faces limitations such as computational overhead \cite{automl_recsys} and reduced dataset-specific optimization. Addressing these limitations is crucial for extending its applicability to more diverse recommendation scenarios.

The main contributions of this position paper are:
\begin{itemize}
    \item We propose \textit{DAReS (Dataset-Agnostic Recommender System)}, which aims to maximize the reusability of recommender system code and minimize the barrier to entry by eliminating the need for dataset-specific configurations for every solution development (Section~\ref{sec:dares}).
    \item We introduce the \textit{Dataset Description Language (DsDL)} (Section~\ref{sec:dsdl}), a structured language that provides a standardized way to describe datasets, enabling autonomous feature engineering and model selection by DAReS.
    \item We define the concept of \emph{level-1 and level-2 automation} for recommender systems, with level-1 focusing on dataset-agnostic but task-specific capabilities, and level-2 representing a fully autonomous system that is both dataset-agnostic and task-agnostic (Section~\ref{sec:leveling}).
\end{itemize}


%% file: content_dares.tex
The \textit{Dataset-Agnostic Recommender System (DAReS)} aims to provide a flexible, reusable solution for handling diverse datasets in recommendation tasks. Central to this approach is the \textit{Dataset Description Language (DsDL)}, which enables the system to interpret various datasets and configure itself accordingly. The DsDL is structured to provide key metadata about the dataset, such as feature types, labels, and task information, all in a standardized format that allows DAReS to generalize across datasets without manual intervention.
\begin{lstlisting}[basicstyle=\ttfamily\scriptsize, frame=single, caption={EBNF Grammar for DsDL}, label={lst:dsdl_grammar}]
DsDL ::= "features" ":" "[" FeatureList "]" [UserID] [ItemID] [Timestamp] 
         [LabelList]
FeatureList ::= Feature { "," Feature }
Feature ::= "{" "col_name" ":" String "," "type" ":" FeatureType "}"
FeatureType ::= "categorical" | "ordinal" | "numeric" | "binary" | "textual" | 
                "url"
UserID ::= "user_id" ":" "{" "col_name" ":" String "}"
ItemID ::= "item_id" ":" "{" "col_name" ":" String "}"
Timestamp ::= "timestamp" ":" "{" "col_name" ":" String "}"
LabelList ::= "label" ":" "[" Label { "," Label } "]"
Label ::= "{" "name" ":" String "," "type" ":" FeatureType "}"
String ::= <any string>
\end{lstlisting}

\subsection{Dataset Description Language (DsDL)}
\label{sec:dsdl}
We provide a formal definition of the DsDL syntax using the Extended Backus-Naur Form (EBNF) \cite{scowen1993generic}, which outlines the structure of the DsDL configuration files in Listing \ref{lst:dsdl_grammar}.

This EBNF grammar describes the key components of DsDL:

\begin{itemize}
    \item \textbf{Features}: A mandatory list of feature descriptions, each consisting of a column name (\texttt{col\_name}) and a type (\texttt{categorical}, \texttt{ordinal}, \texttt{numeric}, \texttt{binary}, \texttt{textual}, or \texttt{url}\footnote{which can be used to fetch rich media, such as images or videos}). 
    \item \textbf{User ID, Item ID, and Timestamp (Optional)}: The \texttt{user\_id}, \texttt{item\_id}, and \texttt{Timestamp} fields are optional. This is because some datasets may not explicitly define these columns or intentionally remain vague about them. In some tasks, such as click-through rate (CTR) prediction \cite{yang2022click}, these columns may not be necessary as the relationships between users and items can be derived from other available features and. However, in other tasks, such as the \textit{Top-N recommendation} task \cite{deshpande2004item}, knowing which column represents the user ID and which represents the item ID is crucial to making accurate recommendations. 
    And in general, timestamp can also be used to produce validation splits. 
    Therefore, while these fields are optional, they are still worth mentioning and treating separately for tasks that require them.
    \item \textbf{Label}: An optional list of labels used for classification or regression tasks, each with a name and a type.
\end{itemize}

It is also worth noting that there might be some other recommender system tasks that require additional specific metadata. We plan to incorporate these additional details gradually, extending the DsDL schema as needed to better support a wide range of recommendation scenarios.

Although the EBNF grammar is used here to formally define the allowable syntax for creating a valid DsDL, the practical implementation could be written in other format for ease of use. 
In Listing \ref{lst:dsdl_example} we provide an example in YAML format.
\begin{lstlisting}[basicstyle=\ttfamily\scriptsize, frame=single, caption={An example of DsDL in YAML format.}, label={lst:dsdl_example}]
DsDL:
  features: [
    { col_name: age, type: numeric },
    { col_name: is_subscriber, type: binary },
    { col_name: product_cat, type: categorical },
    { col_name: product_desc, type: textual },
    { col_name: product_price, type: numeric },
    { col_name: product_satisfaction_level, type: ordinal },
    { col_name: product_image, type: url }
  ]
  user_id: { col_name: usr_id }
  item_id: { col_name: product_id }
  timestamp: { col_name: ts }
  label: [
    { name: purchase_decision, type: binary }
  ]
\end{lstlisting}


\subsection{Autonomous Feature Engineering and Preprocessing}

The DAReS system \emph{may} autonomously handle feature engineering and preprocessing tasks. While including them can significantly enhance the quality of the recommender system, it is still possible to create a functioning DAReS without these autonomous capabilities, albeit with potentially reduced performance. The key processes in this stage include:

\begin{itemize}
    \item \textbf{Feature selection} \cite{erase_ark}: The system selects relevant features based on the descriptions in the DsDL. It identifies the feature types (e.g., numeric, categorical) and determines which features should be used for the task at hand.
    \item \textbf{Missing value handling} \cite{saar2007handling}: Using metadata provided in the DsDL, the system decides how to handle missing values, applying imputation strategies if necessary.
    \item \textbf{Feature transformation} \cite{liu1998feature}: The system applies transformations, such as encoding categorical variables, normalizing numerical features, or extracting embeddings for textual data.
    \item \textbf{Noise removal} \cite{jiang2024more}: The system can automatically detect and handle noisy data based on specified thresholds or through more advanced filtering techniques.
\end{itemize}
Through these data pre-processing steps, DAReS can process datasets automatically to improve performance and streamline setup.
Any automated method for feature engineering can also be incorporated here~\cite{wang2023toward,zhang2023openfe}.

\subsection{Model Selection and Hyperparameter Tuning}

Similarly, DAReS may include model selection and hyperparameter tuning. 
While implementing these automated processes improves the model's performance and ensures that the system adapts optimally to any dataset, it is still possible to create DAReS without these components. 
The steps involved in this process are:
\begin{itemize}
    \item \textbf{Model selection} \cite{elsken2019neural}: selects the appropriate model architecture for the task. 
    \item \textbf{Hyperparameter tuning} \cite{yang2020hyperparameter}: uses automated methods like grid search or Bayesian optimization, the system tunes hyperparameters to maximize performance. Techniques such as early stopping could be used to ensure efficiency.
    \item \textbf{Cross-validation} \cite{schaffer1993selecting}: 
    ensure that the model generalizes well to unseen data, we could use cross-validation to evaluate the performance of the model across different subsets of the training data. This provides an estimate of the model's robustness and helps in avoiding overfitting.
\end{itemize}

\subsection{Model Evaluation}

Once a model has been trained, DAReS may autonomously evaluates its performance to ensure that it meets the desired standards for the specific recommendation task. The evaluation process involves several key components:

\begin{itemize}
    \item \textbf{Performance Metrics}: Depending on the recommendation task, DAReS uses appropriate performance metrics. For example, it may use \textit{AUC-ROC} or \textit{log loss} for CTR prediction, \textit{RMSE} for rating prediction, or \textit{precision@K} and \textit{recall@K} for Top-N recommendation tasks. The use of appropriate metrics ensures that the evaluation aligns with the business goals and the specific requirements of the task.
    \item \textbf{Test Set Evaluation}: If a test set is provided, DAReS uses it to generate predictions and evaluate the final model's performance. This step is crucial for assessing how well the model will perform in a real-world scenario.
\end{itemize}
The model evaluation process ensures that the models produced by DAReS are not only optimized for the training data but also robust and capable of delivering consistent performance on unseen datasets.

%% file: content_comparison.tex
We compare the 
traditional recommender systems 
and DAReS in Table~\ref{tab:comparison}, 
highlighting the key differences in adaptability, human intervention, reusability, reproducibility, task-specific customization and computational overhead. 

\begin{table}[h]
\centering
\caption{Comparison between DAReS and Traditional Recommender Systems}
\label{tab:comparison}
\scriptsize 
\sffamily
\begin{tabular}{p{2.2cm} p{4.5cm} p{5cm}} 
\hline
\textbf{Aspect} & \textbf{Traditional Recommender Systems} & \textbf{DAReS} \\ \hline

\textbf{Adaptability} & Requires significant manual intervention for each new dataset, often tailored for specific use cases. & Automatically adapts to various datasets using the Dataset Description Language (DsDL) without re-engineering. \\ \hline

\textbf{Human Intervention and Expertise} & Requires extensive domain knowledge for feature engineering, model selection, and hyperparameter tuning. & Minimizes human intervention; automates feature engineering, model selection, and hyperparameter tuning, making it accessible to non-experts. \\ \hline

\textbf{Reusability} & Low code reusability due to dataset-specific designs. Significant modifications are needed to adapt to different datasets. & High code reusability enabled by DsDL, allowing the same codebase to work across multiple datasets with minimal or no changes. \\ \hline

\textbf{Reproducibility} & Reproducibility is challenging due to undocumented tweaks and dataset-specific modifications. & Improved reproducibility through standardized dataset descriptions using DsDL, which reduces variability across experiments. \\ \hline

\textbf{Dataset-Specific Optimization} & Capable of deep customization for specific datasets, allowing for highly optimized performance. & Trades off deep customization for generalizability, potentially leading to suboptimal performance in highly specialized tasks. \\ \hline

\textbf{Computational Overhead} & Computationally efficient due to task-specific optimizations and manual configuration focusing on relevant features and models. & Can have significant computational overhead due to automated feature engineering, model selection, and hyperparameter tuning, especially for large-scale datasets. \\ \hline

\end{tabular}
\end{table}

%% file: content_leveling.tex
The development of the Dataset-Agnostic Recommender System (DAReS) can be understood as a progression across different levels of automation, moving from a dataset-agnostic but task-specific system to a fully autonomous, task-agnostic, and dataset-agnostic recommender system. We refer to these levels as \textbf{level-1} and \textbf{level-2} automation, each representing significant milestones in achieving a more generalized and autonomous recommendation framework.

\subsection{Level-1 Automation: Dataset-Agnostic but Task-Specific}

The current definition of DAReS falls under \textbf{level-1 automation}, which is dataset-agnostic but still task-specific. In this phase, DAReS can autonomously adapt to different datasets using the Dataset Description Language (DsDL) without requiring dataset-specific code adjustments. However, the system relies on the task being pre-defined. For instance, tasks such as click-through rate prediction, rating prediction, or Top-N recommendation must be explicitly specified by the user.
This level, however, already provides benefits by reducing manual intervention for data preparation, feature engineering, and model configuration, making DAReS adaptable across diverse datasets, unlike traditional AutoML approaches \cite{he2021automl} which, while automating feature, model and hyperparameter selection, still require dataset-specific configurations and adaptations.

\subsection{Level-2 Automation: Task-Agnostic and Dataset-Agnostic}

The next evolution of DAReS is aimed at achieving \textbf{level-2 automation}, where the system becomes both task-agnostic \emph{and} dataset-agnostic. 
In this phase, DAReS would autonomously determine not only the dataset structure but also infer the appropriate recommendation task based on the provided data. 
This advancement would further reduce the dependency on user input, allowing the system to operate as a fully autonomous recommendation framework. 



%% file: content_conclusion.tex
\textbf{Advantages.}
DAReS provides a reproducible and reusable solution for building recommendation systems across diverse datasets. By leveraging the Dataset Description Language (DsDL), 
it is possible to automate critical tasks such as feature engineering, model selection, and hyperparameter tuning, significantly reducing the need for human expertise and manual intervention. 


\noindent
\textbf{Limitations.}
However, there are certain limitations that need to be addressed, such as high computational overhead and reduced task-specific customization. While these limitations are inherent in the trade-off between generalization and specialization, understanding them is crucial for positioning DAReS effectively across different use cases. 

\noindent
\textbf{Future Work.}
An important future direction for DAReS is to evolve towards becoming a fully \textit{task-agnostic recommender system}. This advancement would involve the system autonomously determining both the recommendation task type and optimal configurations based on the dataset characteristics, without explicit user input. Key steps towards achieving this include enhancing DsDL to provide richer metadata for task inference, and developing more generalized strategies for model selection and adaptive feature engineering. Such enhancements would further reduce the need for manual configuration.

%% file: main.bbl
\begin{thebibliography}{10}

\bibitem{brown2020language}
Tom~B Brown.
\newblock Language models are few-shot learners.
\newblock {\em arXiv preprint arXiv:2005.14165}, 2020.

\bibitem{cantador2011second}
Iv{\'a}n Cantador, Peter Brusilovsky, and Tsvi Kuflik.
\newblock Second workshop on information heterogeneity and fusion in recommender systems (hetrec2011).
\newblock In {\em Proceedings of the fifth ACM conference on Recommender systems}, pages 387--388, 2011.

\bibitem{deshpande2004item}
Mukund Deshpande and George Karypis.
\newblock Item-based top-n recommendation algorithms.
\newblock {\em ACM Transactions on Information Systems (TOIS)}, 22(1):143--177, 2004.

\bibitem{elsken2019neural}
Thomas Elsken, Jan~Hendrik Metzen, and Frank Hutter.
\newblock Neural architecture search: A survey.
\newblock {\em Journal of Machine Learning Research}, 20(55):1--21, 2019.

\bibitem{recwild_2023}
Kim Falk and Morten Arngren.
\newblock Recommenders in the wild - practical evaluation methods.
\newblock In {\em Proceedings of the 17th ACM Conference on Recommender Systems}, RecSys '23, page~1, New York, NY, USA, 2023. Association for Computing Machinery.

\bibitem{repr_dacrema}
Maurizio Ferrari~Dacrema, Simone Boglio, Paolo Cremonesi, and Dietmar Jannach.
\newblock A troubling analysis of reproducibility and progress in recommender systems research.
\newblock {\em ACM Trans. Inf. Syst.}, 39(2), January 2021.

\bibitem{are_dacrema_2019}
Maurizio Ferrari~Dacrema, Paolo Cremonesi, and Dietmar Jannach.
\newblock Are we really making much progress? a worrying analysis of recent neural recommendation approaches.
\newblock In {\em Proceedings of the 13th ACM Conference on Recommender Systems}, RecSys '19, page 101–109, New York, NY, USA, 2019. Association for Computing Machinery.

\bibitem{he2021automl}
Xin He, Kaiyong Zhao, and Xiaowen Chu.
\newblock Automl: A survey of the state-of-the-art.
\newblock {\em Knowledge-based systems}, 212:106622, 2021.

\bibitem{hidasi2023effect}
Bal{\'a}zs Hidasi and {\'A}d{\'a}m~Tibor Czapp.
\newblock The effect of third party implementations on reproducibility.
\newblock In {\em Proceedings of the 17th ACM Conference on Recommender Systems}, pages 272--282, 2023.

\bibitem{erase_ark}
Pengyue Jia, Yejing Wang, Zhaocheng Du, Xiangyu Zhao, Yichao Wang, Bo~Chen, Wanyu Wang, Huifeng Guo, and Ruiming Tang.
\newblock Erase: Benchmarking feature selection methods for deep recommender systems.
\newblock In {\em Proceedings of the 30th ACM SIGKDD Conference on Knowledge Discovery and Data Mining}, KDD '24, page 5194–5205, New York, NY, USA, 2024. Association for Computing Machinery.

\bibitem{jiang2024more}
Gaoxia Jiang, Jia Zhang, Xuefei Bai, Wenjian Wang, and Deyu Meng.
\newblock Which is more effective in label noise cleaning, correction or filtering?
\newblock In {\em Proceedings of the AAAI Conference on Artificial Intelligence}, volume~38, pages 12866--12873, 2024.

\bibitem{automl_position}
Marius Lindauer, Florian Karl, Anne Klier, Julia Moosbauer, Alexander Tornede, Andreas~C Mueller, Frank Hutter, Matthias Feurer, and Bernd Bischl.
\newblock Position: A call to action for a human-centered {A}uto{ML} paradigm.
\newblock In Ruslan Salakhutdinov, Zico Kolter, Katherine Heller, Adrian Weller, Nuria Oliver, Jonathan Scarlett, and Felix Berkenkamp, editors, {\em Proceedings of the 41st International Conference on Machine Learning}, volume 235 of {\em Proceedings of Machine Learning Research}, pages 30566--30584. PMLR, 21--27 Jul 2024.

\bibitem{liu1998feature}
Huan Liu and Hiroshi Motoda.
\newblock Feature transformation and subset selection.
\newblock {\em IEEE Intell Syst Their Appl}, 13(2):26--28, 1998.

\bibitem{liu2022online}
Xianghang Liu, Bart{\l}omiej Twardowski, and Tri~Kurniawan Wijaya.
\newblock {Online Meta-Learning for Model Update Aggregation in Federated Learning for Click-Through Rate Prediction}.
\newblock {\em 28th ACM SIGKDD 2022 Workshop on AdKDD}, 2022.

\bibitem{raffel2020exploring}
Colin Raffel, Noam Shazeer, Adam Roberts, Katherine Lee, Sharan Narang, Michael Matena, Yanqi Zhou, Wei Li, and Peter~J Liu.
\newblock Exploring the limits of transfer learning with a unified text-to-text transformer.
\newblock {\em Journal of machine learning research}, 21(140):1--67, 2020.

\bibitem{ricci2010introduction}
Francesco Ricci, Lior Rokach, and Bracha Shapira.
\newblock Introduction to recommender systems handbook.
\newblock In {\em Recommender systems handbook}, pages 1--35. Springer, 2010.

\bibitem{saar2007handling}
Maytal Saar-Tsechansky and Foster Provost.
\newblock Handling missing values when applying classification models.
\newblock 2007.

\bibitem{schaffer1993selecting}
Cullen Schaffer.
\newblock Selecting a classification method by cross-validation.
\newblock {\em Machine learning}, 13:135--143, 1993.

\bibitem{scowen1993generic}
Roger~S Scowen.
\newblock Generic base standards.
\newblock In {\em Proceedings 1993 Software Engineering Standards Symposium}, pages 25--34. IEEE, 1993.

\bibitem{shao2024rboard}
Xinyang Shao, Edoardo D'Amico, G{\'{a}}bor Fodor, and Tri~Kurniawan Wijaya.
\newblock {RBoard}: {A} {Unified Platform for Reproducible and Reusable Recommender System Benchmarks}.
\newblock {\em CoRR}, abs/2409.05526, 2024.

\bibitem{touvron2023llama}
Hugo Touvron, Thibaut Lavril, Gautier Izacard, Xavier Martinet, Marie-Anne Lachaux, Timoth{\'e}e Lacroix, Baptiste Rozi{\`e}re, Naman Goyal, Eric Hambro, Faisal Azhar, et~al.
\newblock Llama: Open and efficient foundation language models.
\newblock {\em arXiv preprint arXiv:2302.13971}, 2023.

\bibitem{automl_recsys}
Tobias Vente.
\newblock Advancing automation of design decisions in recommender system pipelines.
\newblock In {\em Proceedings of the 17th ACM Conference on Recommender Systems}, RecSys '23, page 1355–1360, New York, NY, USA, 2023. Association for Computing Machinery.

\bibitem{wang2023toward}
Kafeng Wang, Pengyang Wang, and Chengzhong Xu.
\newblock Toward efficient automated feature engineering.
\newblock In {\em 2023 IEEE 39th International Conference on Data Engineering (ICDE)}, pages 1625--1637. IEEE, 2023.

\bibitem{autorec}
Ting-Hsiang Wang, Xia Hu, Haifeng Jin, Qingquan Song, Xiaotian Han, and Zirui Liu.
\newblock Autorec: An automated recommender system.
\newblock In {\em Proceedings of the 14th ACM Conference on Recommender Systems}, RecSys '20, page 582–584, New York, NY, USA, 2020. Association for Computing Machinery.

\bibitem{wu2023mmgef}
Hao Wu, Alejandro Ariza{-}Casabona, Bartlomiej Twardowski, and Tri~Kurniawan Wijaya.
\newblock {MM-GEF:} {Multi-modal representation meet collaborative filtering}.
\newblock {\em CoRR}, abs/2308.07222, 2023.

\bibitem{yang2020hyperparameter}
Li~Yang and Abdallah Shami.
\newblock On hyperparameter optimization of machine learning algorithms: Theory and practice.
\newblock {\em Neurocomputing}, 415:295--316, 2020.

\bibitem{yang2022click}
Yanwu Yang and Panyu Zhai.
\newblock Click-through rate prediction in online advertising: A literature review.
\newblock {\em Information Processing \& Management}, 59(2):102853, 2022.

\bibitem{zhang2023openfe}
Tianping Zhang, Zheyu~Aqa Zhang, Zhiyuan Fan, Haoyan Luo, Fengyuan Liu, Qian Liu, Wei Cao, and Li~Jian.
\newblock Openfe: automated feature generation with expert-level performance.
\newblock In {\em International Conference on Machine Learning}, pages 41880--41901. PMLR, 2023.

\end{thebibliography}
